\documentclass[twocolumn,showpacs,preprintnumbers,draftclsnofoot]{revtex4}
\usepackage{amsmath}
\usepackage{amssymb}
\usepackage{graphicx}
\usepackage{dcolumn}
\usepackage{bm}
\usepackage{amssymb}
\usepackage{graphicx}
\usepackage{dcolumn}
\usepackage{bm}
\begin{document}

\title{ Refractive index sensing based on large negative Goos-H$\ddot{a}$nchen shifts of wavy dielectric grating }
\author{Ma Luo\footnote{Corresponding author:luoma@gpnu.edu.cn}, Feng Wu }%luoma@gpnu.edu.cn,swym231@163.com
\affiliation{School of Optoelectronic Engineering, Guangdong Polytechnic Normal University, Guangzhou 510665, China}

\begin{abstract}

Wavy dielectric grating hosts bound states in the continuum (BICs) at nonzero Bloch wave number. For oblique incident optical field with parameters near to the BICs, the reflectance spectrum exhibits ultra-sharp Fano line shape, and the reflected beam has large negative Goos-H$\ddot{a}$nchen shift, due to excitation of the corresponding quasi-BIC with negative group velocity. Under incidence of Gaussian beam with sizable beam width, the excited quasi-BIC could travel a long distance along the direction of the Goos-H$\ddot{a}$nchen shift, designated as $L_{GH}$, before the energy is completely radiated. If the length between the termination of the wavy shape and the focus of the incident Gaussian beam is smaller than $L_{GH}$, sizable energy flux can be coupled into the waveguide mode of the flat dielectric slab that is connected to the wavy dielectric grating. Measurement of the energy flux of the waveguide mode can sense the variation of the refractive index of the background medium. The proposed sensing scheme can be integrated with waveguide in optical circuit.

\end{abstract}

\maketitle

\section{Introduction}

Refractive index sensing based on optical response is the most efficient mean for detection of environmental changes, such as change of humidity \cite{wang16sen} and gas concentration \cite{Qing21sen}. The critical factor of an efficient sensor is sensitivity, figure of merit (FoM), and the ability to be integrated with photonic circuit. The sensitivity can be enhanced by variant types of optical resonances, such as surface plasmon polaritons (SPPs) and localized surface plasmon resonances (LSPRs) \cite{SPP001,SPP002,SPP003,SPP004,SPP005,SPP006,SPP007,SPP008,SPP009,SPP010,SPP011}. Although SPPs and LSPRs in metallic nano-structures can highly localize the optical field, and then enhance the sensitivity, the absorption losses at visible and near-infrared wavelengths weaken the Q factor of the resonant mode \cite{SPP013}. Thus, the FoMs of sensors based on SPPs and LSPRs are not significantly high. Sensors based on fully dielectric structures, such as resonant nanocavities in 2D photonic crystals \cite{Qing21sen} or dielectric gratings \cite{FengWuSensing}, have ultra-low absorption losses, so that the Q factor can be high, and then the FoM can be increased.

Bound state in the continuum (BIC) is a type of localized eigen modes, whose resonant frequency immerses into continuous spectrum of radiative optical modes \cite{Sadreev21,ChiaWeiHsu16,Marinica08,Bulgakov08,Plotnik11,BoZhen14,YiYang14,TaoXu08,Jeongwon2012,Koshelev18}. Because the BIC does not couple with the radiative mode, incident optical field cannot excite the BIC. By tuning the structural parameter, the BIC can be transferred into quasi-BIC with ultra-high but finite Q factor, which couples with the radiative mode. Under incidence of plane wave with fixed wavelength, the angular reflectance spectrum (i.e., reflectance versus incident angle) of the dielectric structure with quasi-BIC exhibits ultra-narrow Fano line shape, so that the refractive index sensor based on quasi-BIC could have large FoM \cite{FengWuSensing}. For a realistic sensor, the incident optical field is Gaussian beam instead of plane wave, so that the measured angular reflectance spectrum is the convolution between the Fano line shape and the angular spectrum of the incident Gaussian beam. If the beam width of the Gaussian beam is small, the width of the incident angular spectrum is larger than the width of the Fano line shape, so that the width of the measured angular reflectance spectrum is larger than that of the Fano line shape. Thus, the FoM is decreased. On the other hand, if the beam width of the incident Gaussian beam is ultra-large, the FoM is preserved, but the measurement of the reflected or transmitted power requires detector with large cross section, which increases the difficulty of integrating the sensor with photonic circuit.

Another feature of quasi-BIC is the enhancement of the Goos-H$\ddot{a}$nchen shifts \cite{Goos47,Artmann48}, which are lateral shifts of the reflected and transmitted beams from the incident beam. The Goos-H$\ddot{a}$nchen shifts can also be enhanced by the presence of other types of resonant microstructures \cite{Kaiser96}, such as surface plasmon resonators \cite{XYin04}, metal cladding waveguides \cite{YWang17}, epsilon-near-zero metamaterial slabs \cite{YXu15,JWen17}, and photonic crystal surface structures \cite{LWang06,JWu20,YChen16,WZhen20}.

%The positive Goos-H$\ddot{a}$nchen shift is more useful for the practical application that need to manipulate the propagation of light, such as sensing and wavelength division (de)multiplexing \cite{XWang13,XWang16,Sattari16}. The negative Goos-H$\ddot{a}$nchen shift is more useful for the practical application that near to confine light, such as light storage \cite{Tsakmakidis07,RYang16,THuang19}.

\begin{figure*}[tbp]
\scalebox{0.53}{\includegraphics{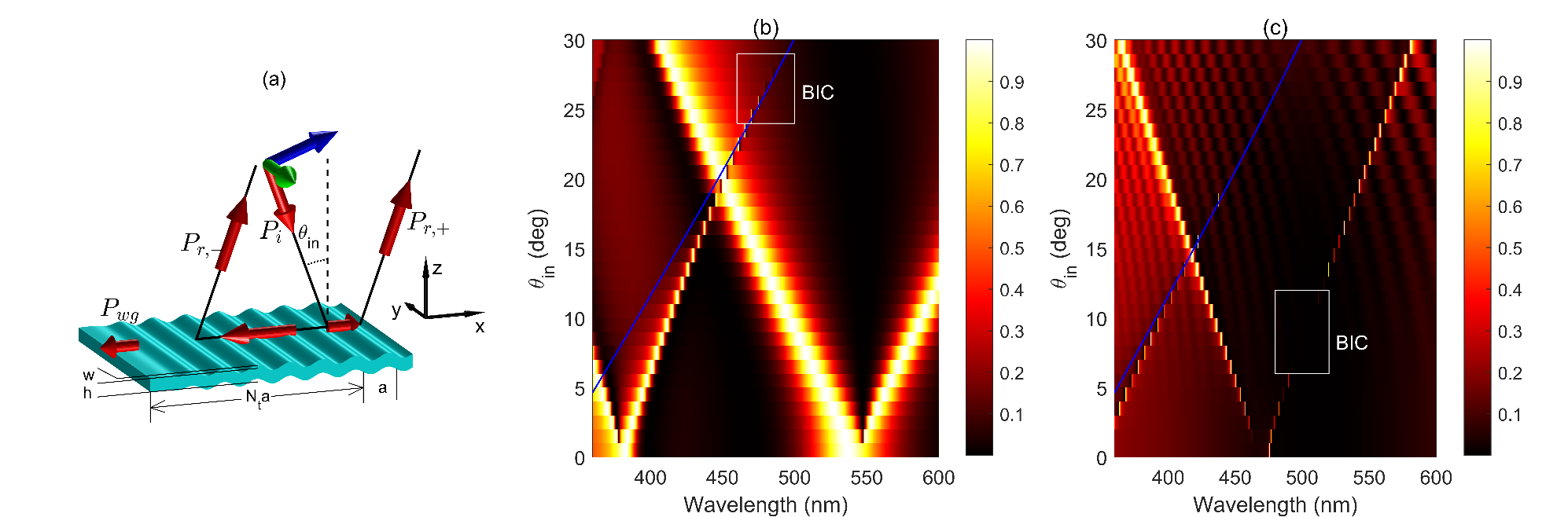}}% Here is how to import EPS art
\caption{ (a) The structure of the wavy dielectric grating, and the scheme of sensing based on in-plane oblique incident Gaussian beam. The red arrows indicate the main patch of the incident optical beam. The power flux of the incident beam is designated as $P_{i}$. The power flux of the reflected beams with positive and negative Goos-H$\ddot{a}$nchen shifts are designated as $P_{r,+}$ and $P_{r,-}$, respectively. The power flux that couple into the flat waveguide adjacent to the left termination of the wavy shape is designated as $P_{wg}$. The green and blue arrows indicate the linearly polarized directions of the s- and p-polarized incident wave, respectively. (b-c) The reflectance spectrum of in-plane oblique incident plane wave versus wavelength and incident angle. The blue line marks the relation $\frac{2\pi}{a}-k_{x}=k_{0}$, which separates the regimes that can and cannot have second order reflection at angle $-\sin^{-1}(\lambda/a-\sin\theta_{in})$. The spectra in (b) and (c) are for the incidence of s-polarized and p-polarized plane wave, respectively. The BICs in (b) and (c) are marked by the white rectangles.   }
\label{figure_inplaneSpec}
\end{figure*}

In this paper, we propose a refractive index sensor based on large negative Goos-H$\ddot{a}$nchen shifts of the quasi-BICs. The sensing signal is the energy flux of the waveguide mode, so that the sensor can be integrated with photonic circuit. The structure consists of a dielectric slab on x-y plane, as shown in Fig. \ref{figure_inplaneSpec}(a). Within a finite region $x\in[-N_{t}a,N_{t}a]$, the dielectric slab has wavy structure along x direction, with $a$ being the period of the wavy structure and $N_{t}$ characterizing the size of the grating. In the region with $|x|>N_{t}a$, the dielectric slab is flat. In the region with $(N_{t}-3)a<|x|<N_{t}a$, the magnitude of the wavy shape smoothly decreases to zero. The wavy dielectric grating within the region with $|x|<(N_{t}-3)a$ host BICs at nonzero Bloch wave number, so that in-plane oblique incident of optical field could excited the corresponding quasi-BICs \cite{maluo22}. According to the stationary phase method \cite{Artmann48}, the Goos-H$\ddot{a}$nchen shifts assisted by the quasi-BICs are large \cite{fengwu19,fengwu21}. The quasi-BICs of wavy dielectric grating have negative group velocity, so that the Goos-H$\ddot{a}$nchen shifts is negative. Assuming that the focus of the incident Gaussian beam is at $x=0$, the incident field excites wave packet consisted of the quasi-BICs with energy flux along $-\hat{x}$ direction. As the excited wave packet travels to the region with $x<0$, the optical field couples into the radiative modes, which form reflected and transmitted beams with negative Goos-H$\ddot{a}$nchen shifts, as designated by the red arrow $P_{r,-}$ in Fig. \ref{figure_inplaneSpec}(a). If $N_{t}a$ is smaller than sum of the Goos-H$\ddot{a}$nchen shift and the reflected beam width, the energy of the wave packet is not completely radiated before reaching $x=-N_{t}a$, but is partially coupled into the waveguide mode of the flat dielectric slab in the region with $x<-N_{t}a$, as designated by the red arrow $P_{wg}$ in Fig. \ref{figure_inplaneSpec}(a). In our proposed systems, the magnitude of the negative Goos-H$\ddot{a}$nchen shift is larger than the beam width of the incident Gaussian beam, so that the termination of the wavy shape at $x=-N_{t}a$ does not influence the interaction between the incident Gaussian beam and the wavy dielectric grating. Thus, the large negative Goos-H$\ddot{a}$nchen shift is preserved, and then $P_{wg}$ is sizable. Measurement of $P_{wg}$ by integrated photonic detector can be applied as sensing signal of the refractive index in the background medium. Comparing to the previous work in Ref. \cite{fengwu19,fengwu21}, we applied Gaussian beam in real space to excite the quasi-BIC with large negative Goos-H$\ddot{a}$nchen shift, so that the sensing signal is the energy flux in the integrated waveguide, instead of the reflected (transmitted) plane wave or Gaussian beam.

The remainder of this paper is organized as follows. In Sec. II, the reflectance spectrum, sensitivity and FoM under plane wave incidence to the structure with $N_{t}=\infty$ are investigated. In Sec. III, the Goos-H$\ddot{a}$nchen shifts of two selected quasi-BICs are calculated by the stationary phase method; the Goos-H$\ddot{a}$nchen shifts in real space for the structure with finite $N_{t}$ under incidence of Gaussian beam with finite beam width are calculated; the sensing scheme based on measurement of the tunneling energy into the flat waveguide is discussed. In Sec. IV, conclusions are presented.

\section{Optical respond under plane wave incidence }

The structure of wavy dielectric grating is consisted of periodically corrugated dielectric slab with thickness being $h$, as shown in Fig. \ref{figure_inplaneSpec}(a). Assuming that the slab lays on the x-y plane, the shapes of the top and bottom surface of the dielectric slab within the region $|x|<N_{t}a$ are given by the functions $z=w\sin(2\pi x/a)f(x)\pm h/2$, where $w$ is the magnitude of the corrugation and $f(x)=(1+\tanh[2(x/a-1.5+N_{t})])(1-\tanh[2(x/a+1.5-N_{t})])/4$ is a smoothed ladder function at the two terminations of the wavy shape. The wavy shape can be fabricated by applying the femtosecond laser direct-writing technique \cite{fsdwtech21}. In this section, the scattering of oblique incident plane wave, whose plane of incident is the x-z plane, are studied. The wave equations of the electric field ($E_{y}$) and magnetic field ($H_{y}$) for the s-polarized and p-polarized incident wave are numerical solved by spectral element method (SEM) \cite{SEM1,SEM2,SEM3,SEM4,SEM5,SEM6}, respectively. In the numerical simulation, the structural parameters are $a=333$ nm, $h=134$ nm, $w=30$ nm. $N_{t}$ is infinitely large, so that periodic boundary condition of one period can be applied to simulate the plane wave scattering. The refractive index of the dielectric slab is $n=2$. The refractive index of the background medium above the wavy dielectric grating is 1, and that below the wavy dielectric grating is $n_{sen}$. We firstly discuss the case with $n_{sen}=1$, and then study the sensitivity of the optical response to the changing of $n_{sen}$.

\subsection{Reflectance spectrum}

\begin{figure}[tbp]
\scalebox{0.68}{\includegraphics{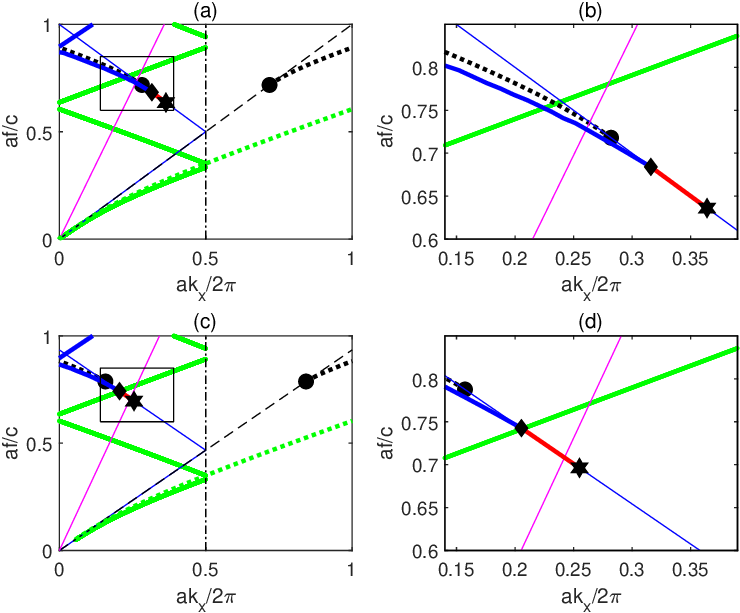}}% Here is how to import EPS art
\caption{ The band structure of the s-polarized leaky resonant modes with $n_{sen}=1$ and $n_{sen}=1.07$ in (a,b) and (c,d), respectively. The panels (b) and (d) are zoom in views of the panels (a) and (b) within the rectangular regime, respectively. The dashed thin black line is the light cone of the lower half space given by $k_{x}=2\pi fc/n_{sen}$. The solid thin blue line is the folded light cone within the first Brillouin zone. The thick dotted green (black) lines below the light cone are the first (second) waveguide dispersion of a flat dielectric slab with $w=0$ nm. The cutoff mode of the second waveguide mode is marked by the circle black dot. The second waveguide mode and its cutoff mode is formally folded into the first Brillouin zone. The thick green line and the thick blue-red line are the band structure of the first and second leaky resonant modes of the wavy dielectric grating, which is originated from the first and second waveguide dispersion, respectively. The band structure of the second leaky resonant mode cross the folded light cone at the diamond dot, and terminates at the hexagram dot. The solid thin magenta line is the light line given by $k_{x}=(2\pi f/c)\sin{\theta_{in}}$.    }
\label{figure_SensingScheme}
\end{figure}

\begin{figure}[tbp]
\scalebox{0.48}{\includegraphics{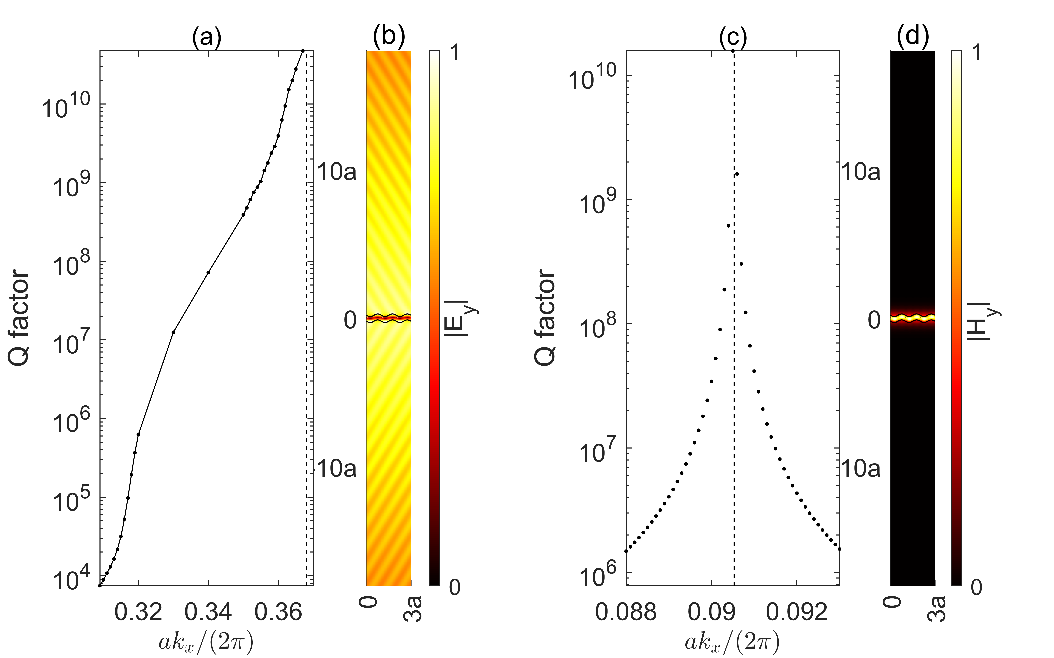}}% Here is how to import EPS art
\caption{ The Q factor versus $k_{x}$ of the quasi-BIC of the s-polarization in (a), and the p-polarization in (c). The field pattern of $|E_{y}|$ and $|H_{y}|$ of the s-polarized and p-polarized BICs are plotted in panels (b) and (c), respectively.    }
\label{figure_Qfactor}
\end{figure}

The numerical results of reflectance spectrum for s-polarized and p-polarized incident plane wave in the parameter space of wavelength (designated as $\lambda$) and incident angle (designated as $\theta_{in}$) are plotted in Figs. \ref{figure_inplaneSpec}(b) and (c), respectively. Within the white rectangle in Figs. \ref{figure_inplaneSpec}(b) and (c), the line width of the reflectance spectrum versus wavelength approaches zero, which imply existence of a BIC. The thin blue line corresponds to the folded light cone given by $\frac{2\pi}{a}-k_{x}=k_{0}$, with $k_{0}=2\pi/\lambda$ and $k_{x}=\sin\theta_{in}k_{0}$.

For s-polarized wave, four bands of leaky resonant modes are identified in Fig. \ref{figure_inplaneSpec}(b), which are correspond to the band structure above the light cone. In the absence of the wavy shape (i.e., $w=0$), the flat dielectric slab possesses waveguide modes, whose dispersion relations are plotted as dotted lines in Fig. \ref{figure_SensingScheme}(a). The green and black dotted lines are for the first and second waveguide modes, respectively. By imposing the discrete periodical boundary condition, the dispersion relation of the waveguide modes are folded into the first Brillioun zone, which form a band structure of the waveguide modes. Specifically, the cutoff mode of the second waveguide band is folded into the first Brillioun zone, so that the mode is at the folded light cone given by $\frac{2\pi}{a}-k_{x}=k_{0}$, as shown by the black circle dot in Fig. \ref{figure_SensingScheme}(a,b). As $w$ become nonzero, the perturbation modifies the band structure of the waveguide modes, and transfers the modes above the light cone into leaky resonant modes. The band structures of the leaky resonant modes, which are plotted as thick solid lines in Fig. \ref{figure_SensingScheme}(a,b), can be obtained by solving the eigenvalue problem of the unit cell with radiative boundary conditions at the top and bottom surfaces. The band crossing of the waveguide modes at $k_{x}=0$ are transferred into avoid crossing, with one of the two modes being symmetry-protected BIC with real eigenvalue. The perturbation induces coupling between the waveguide mode near to the cutoff frequency of the second waveguide band and the propagating mode at the light cone, which modifies the band structure near to the cutoff frequency of the second waveguide mode. The band structure becomes lower than the dispersion of the waveguide mode, as shown by the thick blue line in Fig. \ref{figure_SensingScheme}(a,b). At the diamond point with $ak_{x}/(2\pi)=0.316$, the band structure crosses the folded light cone. As $k_{x}$ further increases, the band structure further extend with frequency slightly above the folded light cone, as shown by the thick red line in Fig. \ref{figure_SensingScheme}(a,b). The band structure terminates at the hexagram point with wave number being $ak_{x,cBIC}/(2\pi)=0.3681$. The Q factors of the leaky resonant modes along the band structure (from the thick blue line to the thick red line) are plotted in Fig. \ref{figure_Qfactor}(a). As $k_{x}$ approaches $k_{x,cBIC}$, the Q factor increases and approaches infinite, which implies that the resonant mode at $k_{x,cBIC}$ is a BIC. We designated the BIC as cutoff-BIC, because the BIC is originated from the cutoff waveguide mode. The field pattern of a quasi-BIC at $ak_{x}/(2\pi)=0.36$ is plotted in Fig. \ref{figure_Qfactor}(b), which is highly non-localized. Because the field pattern of the BIC and the corresponding quasi-BICs is highly non-localized, a unit cell with large size along z direction is required in the numerical calculation to obtain accurate result. For a quasi-BIC at $k_{x}$, the field pattern can be decomposed into superposition of serial of Fourier modes with Bloch wave number being $k_{x}-N_{w}2\pi/a$ (with $N_{w}$ being integer). The Fourier mode with $k_{x}-2\pi/a$ is near to the light cone, whose evanescent field is nearly non-localized. Because the quasi-BIC is originated from the coupling between the propagating mode at the light cone and the waveguide mode, the superposition coefficient of the Fourier mode with wave number being $k_{x}-2\pi/a$ is sizable. Thus, the field pattern is highly non-localized. As $k_{x}$ approaches $k_{x,cBIC}$, the Q factor of the quasi-BIC approaches infinite with a varying pace, i.e., $Q\propto1/|k_{x}-k_{x,cBIC}|^{p}$ with a varying index $p$. This phenomenon might be due to the highly delocalization of the field pattern.

Under the plane wave incidence, the reflectance spectrum has resonant peak near to the crossing point between the band structure and the light line given by $k_{x}=k_{0}\sin\theta_{in}$. As $\theta_{in,BIC}>28^{o}$, the resonant peak become ultra-sharp, because the Q factor of the leaky resonant mode at the crossing point is larger than $10^{6}$. The group velocity of the quasi-BICs is negative, because the slope of the corresponding band structure is negative.

For p-polarized wave, three bands of leaky resonant modes are identified in Fig. \ref{figure_inplaneSpec}(c). The band structure near to the blue line is also consisted of the quasi-BICs corresponding to the cutoff-BIC, which has similar feature as that of the s-polarized wave. Another BIC is identified with wavelength being $\lambda_{BIC}=506$ nm and incident angle being $\theta_{in,BIC}=9^{o}$, which is marked by the white rectangular. The Q factors of the corresponding quasi-BICs versus wave number are plotted in Fig. \ref{figure_Qfactor}(c), which satisfies the condition $Q\propto1/|k_{x}-k_{x,BIC}|^{2}$. Thus, the BIC is a typical accidental-BIC. The field pattern of the accidental-BIC is plotted in Fig. \ref{figure_Qfactor}(d), which is highly localized within the wavy dielectric slab. The group velocity of the corresponding quasi-BICs are also negative.

\subsection{Sensitivity and FoM under plane wave incidence}

\begin{figure}[tbp]
\scalebox{0.62}{\includegraphics{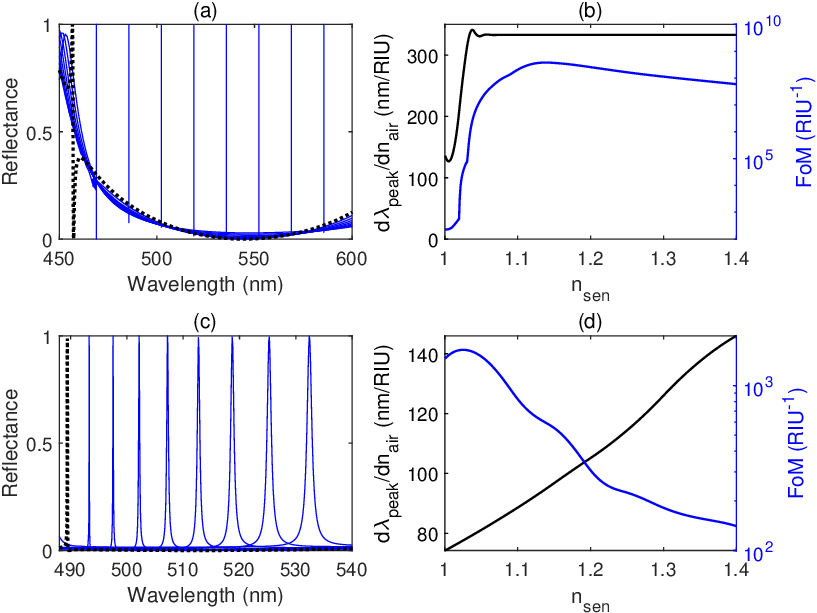}}% Here is how to import EPS art
\caption{ The left column are numerical results of reflectance spectrum of the incident plane wave.  In each figure, the black dashed curve is for the case with $n_{sen}=1$; the blue solid curves with increasing peak wavelength are for the cases with increasing $n_{sen}$ with incremental step being 0.05. The right column are sensitivity and FoM versus $n_{sen}$, which are plotted as black and blue lines, corresponding to the left and right y-axis, respectively. The top and bottom rows are numerical results for s-polarization and p-polarization, with $\theta_{in}$ being 21$^{o}$ and 5$^{o}$, respectively.   }
\label{figure_planeSen}
\end{figure}

Changing $n_{sen}$ changes the resonant frequency of the BICs, which in turn changes the reflectance spectrum of the corresponding quasi-BICs. In another words, the wavelength of the reflectance peak of the quasi-BIC is sensitive to the change of refractive index. The sensitivity of the reflectance peak to $n_{sen}$ is defined as \cite{sensitivity1,sensitivity2}
\begin{equation}
S=\frac{d\lambda_{peak}}{dn_{sen}}
\end{equation}
, and the FoM is defined as
\begin{equation}
F=\frac{S}{\Delta\lambda_{peak}}
\end{equation}
, where $\lambda_{peak}$ is the wavelength at the peak of the reflectance, and $\Delta\lambda_{peak}$ is the full-width half-maximum (FWHM) of the peak.

For s-polarized wave, as $n_{sen}$ increases, the slope of the light cone of the lower half space decreases, and the frequency of the cutoff-BIC increases, as shown in Fig. \ref{figure_SensingScheme}(c,d). The band structure of the quasi-BIC corresponding to the cutoff-BIC shifts to the regime with larger frequency and smaller $k_{x}$. Consequently, the frequency of the crossing point between the light line with a fixed incident angle $\theta_{in}$ and the band structure decreases. When $n_{sen}$ is slightly larger than 1, the crossing point is at the blue section of the band structure, which correspond to the quasi-BIC with moderate Q factor. In this case, the resonant peak of the reflectance spectrum have sizable FWHM, so that the FoM is small. When the increase of $n_{sen}$ is large enough, the crossing point is at the red section of the band structure, which is very near to the folded light cone. Thus, the central wavelength of the resonant peak can be approximately given by the crossing between the light line and the folded light cone, i.e., $\lambda_{peak}=a(n_{sen}+\sin\theta_{in})$. The sensitivity is $S=a$ RIU$^{-1}$=333 nm/RIU. The quasi-BIC at the red section of the band structure has large Q factor, so that the FWHM of the resonant peak is small, which result in large FoM. The numerical result of the reflectance spectrums versus incident wavelength with varying $n_{sen}$ are plotted in Fig. \ref{figure_planeSen}(a). As $n_{sen}$ increases from 1 to 1.06, the sensitivity sharply increase from $115$ nm/RIU to $333$ nm/RIU; as $n_{sen}$ further increases to 1.4, the sensitivity remain being $333$ nm/RIU, as shown by the black line in Fig. \ref{figure_planeSen}(b). As $n_{sen}$ increases from 1 to 1.13, the FoM sharply increase from $2.3\times10^{2}$ RIU$^{-1}$ to $3.7\times10^{8}$ RIU$^{-1}$; as $n_{sen}$ further increases to 1.4, the FoM slowly decreases to $6\times10^{7}$ RIU$^{-1}$, as shown by the blue line in Fig. \ref{figure_planeSen}(b). As a result, the sensing range within $n_{sen}\in[1,1.4]$ has large FoM, which can be applied for sensing of refractive index of gas, liquid, or bio-solution.

For p-polarized wave with $n_{sen}=1$, the accident-BIC appears at $\theta_{in}=9^{o}$ and $\lambda=506$ nm. If $\theta_{in}$ is fixed to be 5$^{o}$, which is near to $9^{o}$, the quasi-BIC with moderate Q factor is excited, as shown by the black dashed line in Fig. \ref{figure_planeSen}(c). As $n_{sen}$ increases, the peak shift to higher wavelength with linear incremental shift, so that the sensitivity is linearly dependent on $n_{sen}$, as shown by the black line in Fig. \ref{figure_planeSen}(d). As $n_{sen}$ increases, the FWHM slowly increases, so that the FoM slowly decreases from $1.6\times10^{3}$ RIU$^{-1}$ to $1.4\times10^{2}$ RIU$^{-1}$, as shown by the blue line in Fig. \ref{figure_planeSen}(d). Because the field pattern of the p-polarized quasi-BICs is localized within the dielectric slab, the sensitivity is smaller than that of the s-polarized quasi-BICs.

\section{Optical respond under Gaussian beam incidence}

\begin{figure*}[tbp]
\scalebox{0.54}{\includegraphics{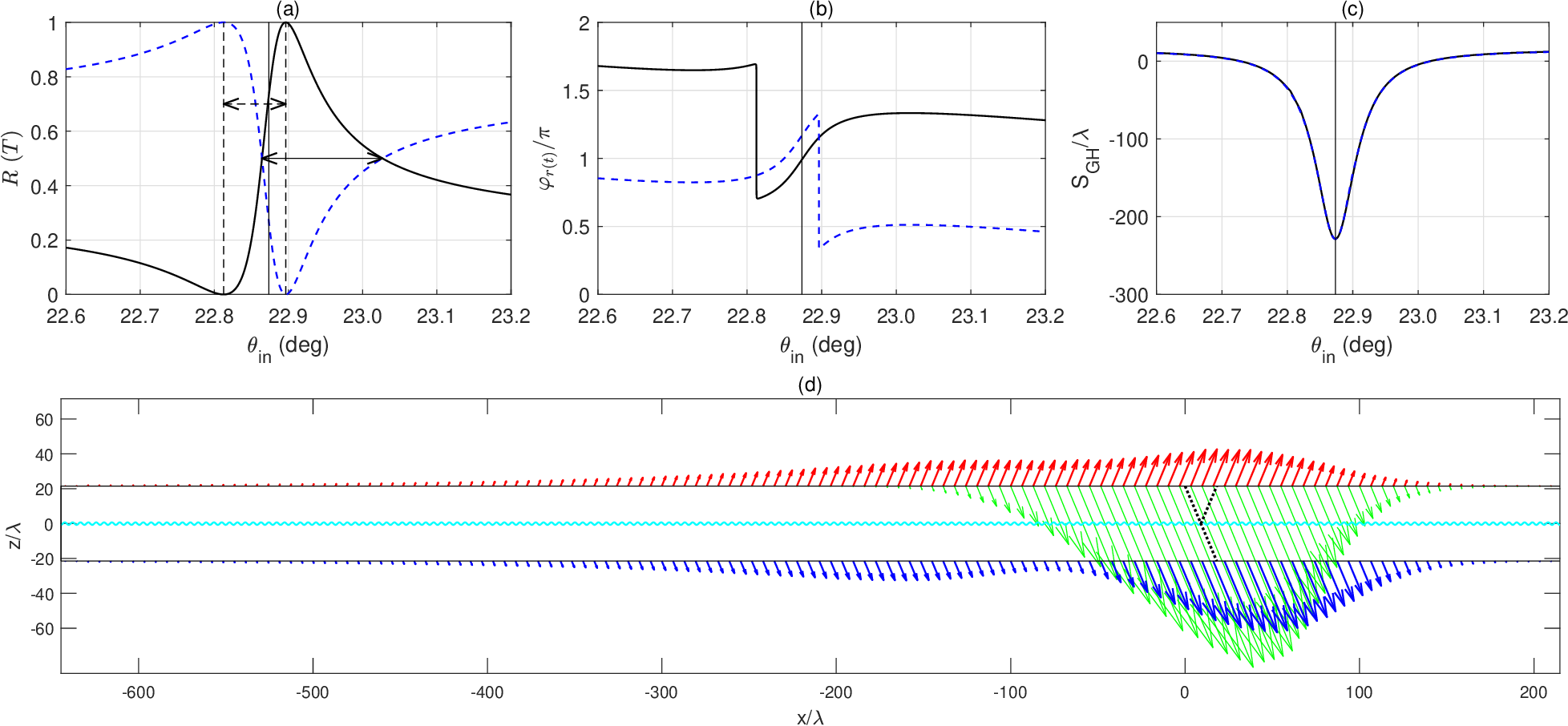}}% Here is how to import EPS art
\caption{ For s-polarization, the reflectance (transmittance) angular spectra for quasi-BIC with $\lambda=465$ nm are plotted as black solid (blue dashed) line in (a). (b) Corresponding reflection (transmission) phase angular spectra. (c) Corresponding Goos-H$\ddot{a}$nchen shift angular spectra. The incident angle with maximum Goos-H$\ddot{a}$nchen shift are marked by vertical thin solid line in (a-c). The incident angles with reflectance being zero and one are marked by vertical thin dashed lines in (a). The FWHM and Fano line-width are marked by horizontal solid and dashed double arrows, respectively. (d) For Gaussian beam with $\lambda=465$ nm, $\theta_{Inc}=22.8735^{o}$, and $w_{0}=100\lambda$, the spatial distribution of time-averaged Poynting vector of the incident and the reflected fields along the horizontal observation plane that is $30a$ above the wavy top surface are plotted as green and red arrows; that of the transmitted field along the horizontal observation plane that is $30a$ below the wavy top surface are plotted as blue arrows. The axis of the incident and normal reflected beam are plotted as thick dashed line. The horizontal cyan wavy curve indicate the location of the wavy dielectric grating.   }
\label{figure_GSshiftS}
\end{figure*}

In this section, we consider the incident Gaussian beam with finite beam width in x-z plane, but infinite beam width along y axis (i.e. the field is uniform along y direction). For the Gaussian beam with fixed wavelength and beam width, the field profile in the x-z plane is given as
\begin{equation}
\mathbf{E}_{0}(\rho,\eta)=\mathbf{E}_{g0}\sqrt{\frac{w_{0}}{w(\xi)}}e^{-\frac{\rho^{2}}{w^{2}(\xi)}+ik_{0}\xi+ik_{0}\frac{\rho^{2}}{2R(\xi)}-\frac{i}{2}\eta(\xi)} \label{gaussianBeam}
\end{equation}
, where $\xi=(\mathbf{r}-\mathbf{r}_{p0})\cdot\mathbf{k}_{Inc}$, $\rho=|\mathbf{r}-\mathbf{r}_{p0}-\xi\mathbf{k}_{Inc}|$, $w(\xi)=w_{0}\sqrt{1+(\xi/z_{0})^{2}}$, $R(\xi)=\xi[1+(z_{0}/\xi)^{2}]$, $\eta(\xi)=\tan^{-1}(\xi/z_{0})$, $z_{0}=\frac{k_{0}w_{0}^{2}}{2}$, with $w_{0}$ being the beam width at the beam waist, $\mathbf{r}_{p0}$ being the location of the focus point, $\mathbf{k}_{Inc}=-\cos\theta_{Inc}\hat{z}+\sin\theta_{Inc}\hat{x}$ being the unit vector along the incident direction $\theta_{Inc}$, and $\mathbf{E}_{g0}=E_{y0}\hat{y}$ with $E_{y0}$ being the amplitude of the Gaussian beam. The focus of the beam is at the top surface of the wavy dielectric slab, given as $\mathbf{r}_{p0}=\hat{x}(30a-h/2)\tan\theta_{Inc}+\hat{z}h/2$. Performing Fourier transformation to the field profile near to $\mathbf{r}_{p0}$, the Gaussian beam can be expanded into superposition of serial of plane waves with the same wavelength and varying incident angle. The incident angular spectrum, is given by Gaussian function $\Theta(\theta)=e^{-[(\theta-\theta_{Inc})/\Delta\theta_{in}]^{2}/2}$, with $\Delta\theta_{in}=\sin^{-1}[\lambda/(\sqrt{2}\pi w_{0})]$ being the divergence angle. Thus, the scattering of the Gaussian beam can be decomposed into scattering of a serial of plane waves with the same wavelength, varying incident angle and varying amplitude. According to the stationary phase method, the Goos-H$\ddot{a}$nchen shifts for the Gaussian beam with $\Delta\theta_{in}$ being infinitely small is given as  \cite{Artmann48}
\begin{equation}
S_{GH}=-\frac{\lambda}{2\pi}\frac{\partial\varphi_{r(t)}}{\partial\theta_{in}}
\end{equation}
, where $\varphi_{r(t)}$ is the phase of the reflected (transmitted) wave. $\varphi_{r(t)}$ can be extracted from the field pattern at the top (bottom) boundary of the computational domain in the simulation of plane wave incidence by applying SEM. On the other hand, $\varphi_{r(t)}$ can also be calculated by applying multiple-layer rigorous coupled wave analysis (RCWA) method \cite{Moharam95RCWA}. We have confirmed that the numerical results of $\phi_{r(t)}$ given by the two methods are identical.

\begin{figure*}[tbp]
\scalebox{0.54}{\includegraphics{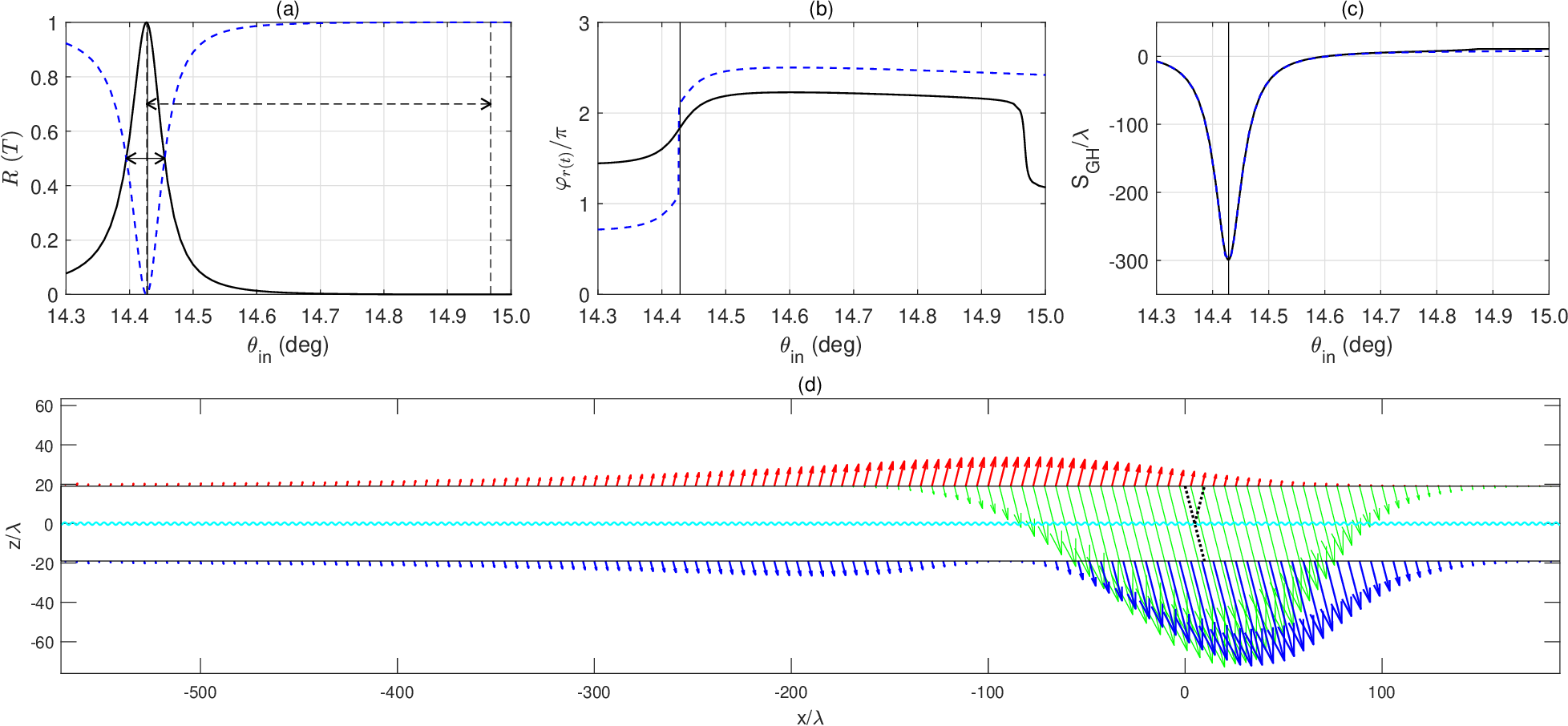}}% Here is how to import EPS art
\caption{ The same as Fig. \ref{figure_GSshiftS}, except that the incident field is p-polarization with $\lambda=525$ nm. In (d), $\theta_{Inc}=14.4285^{o}$.  }
\label{figure_GSshiftP}
\end{figure*}

For realistic structure, $N_{t}$ is finite. In the numerical simulation, the flat dielectric slab in the region $|x|>N_{t}a$ extends into the perfectly match layer (PML) region. Thus, the calculated model simulated the structure of wavy dielectric grating with finite number of periods, whose left and right terminations are connected to infinitely long flat dielectric slab in the background.

\subsection{s-polarization}

For s-polarization, the quasi-BIC at $\lambda=465$ nm is simulated. The reflectance (transmittance) angular spectra is plotted in Fig. \ref{figure_GSshiftS}(a), which exhibit the Fano line shape. The line width of the reflectance peak is defined in two ways. The Fano line-width is defined as difference between $\theta_{in}$ with reflectance being zero and one, which is $0.08373^{o}$. The FWHM is defined as difference between $\theta_{in}$ with reflectance being $0.5$, which is $0.1625^{o}$. The FWHM is larger than the Fano line-width, so that the line shape is highly asymmetric. The reflection (transmission) phase angular spectra and the corresponding Goos-H$\ddot{a}$nchen shift angular spectra are plotted in Fig. \ref{figure_GSshiftS}(b) and (c), respectively. At $\theta_{in}$ with reflectance (transmittance) being zero, the reflection (transmission) phase has a jump of $\pi$, so that the derivatives of the phase against $\theta_{in}$, and then the corresponding Goos-H$\ddot{a}$nchen shift at the corresponding $\theta_{in}$, is not well defined. As $\theta_{in}\in[22.73^{o},23.01^{o}]$, the Goos-H$\ddot{a}$nchen shift is negative. The maximum magnitude of the negative Goos-H$\ddot{a}$nchen shift is $229\lambda$ at $\theta_{in}=22.8735^{o}$. The corresponding $\theta_{in}$ is marked by vertical solid thin line in Fig. \ref{figure_GSshiftS}(a-c), which shows that the corresponding reflectance is $0.732$.

In the simulation of Gaussian beam incident with finite beam width, $\theta_{Inc}$ equates to $\theta_{in}$ with maximum magnitude of the negative Goos-H$\ddot{a}$nchen shift. For Gaussian beam with $w_{0}=100\lambda$, the divergence angle, i.e., $2\Delta\theta_{in}=0.2579^{o}$, is larger than the Fano line-width and the FWHM. The angular spectra of the reflected and transmitted beams equate to multiplication between the incident angular spectrum and the reflected and transmitted angular spectrum, respectively. Within the angular spectra of the reflected and transmitted beams, the amplitude in large range of $\theta_{in}$ with $S_{GH}>0$ is sizable. As a result, the reflected beam consists of two parallel beams, one of which has small positive Goos-H$\ddot{a}$nchen shift, and the another one of which has large negative Goos-H$\ddot{a}$nchen shift, as schematically indicated by the beams with power flux being $P_{r,+}$ and $P_{r,-}$ in Fig. \ref{figure_inplaneSpec}(a), respectively. Similarly, the transmitted beam also consists of two parallel beams. Because $N_{t}$ is finite, before being radiated to the reflected (transmitted) beam, part of the energy of the quasi-BIC is coupled into the left flat dielectric slab with energy flux being $P_{wg}$, as schematically indicated in Fig. \ref{figure_inplaneSpec}(a). The calculated result of a system with $N_{t}=900$ is plotted in Fig. \ref{figure_GSshiftS}(d), which shows that reflected (transmitted) beam does consist of overlapping of two parallel beams with opposite Goos-H$\ddot{a}$nchen shifts. Because $\theta_{Inc}$ is at the highly asymmetric part of the Fano line shape, the reflected beam with large negative Goos-H$\ddot{a}$nchen shift has large beam width and large overlap with the another reflected beam with small positive Goos-H$\ddot{a}$nchen shift. Because $N_{t}$ is large enough, $P_{wg}$ is negligible.

The numerical Goos-H$\ddot{a}$nchen shift of the reflected and transmitted beam can be approximated as the distance between the coordinate of the regular beam center and the coordinate with local maximum energy flux at an observation plane outside of the wavy dielectric slab. By extracting the numerical result in Fig. \ref{figure_GSshiftS}(d), for the reflected beam, the Goos-H$\ddot{a}$nchen shift of the beam with smaller energy flux is $-131\lambda$, and that of the another beam with larger energy flux is nearly zero. For the transmitted beam, the Goos-H$\ddot{a}$nchen shift of the beam with smaller energy flux is $-182\lambda$, and that of the another beam with larger energy flux is $17\lambda$. The numerical results are close to the values of $S_{GH}$ given by the stationary phase method in Fig. \ref{figure_GSshiftS}(c).

\subsection{p-polarization}

For p-polarization, we simulation the quasi-BIC at $\lambda=525$ nm. The same set of numerical results as those for the s-polarization are plotted for the p-polarization in Fig. \ref{figure_GSshiftP}. The Fano line shape of the p-polarized quasi-BIC is highly different from that of the s-polarized quasi-BIC. The Fano line-width ($0.5410^{o}$) is much larger than the FWHM ($0.06000^{o}$). Thus, the Fano line shape is nearly symmetric about the $\theta_{in}$ with reflectance being one. The $\theta_{in}$ with maximum magnitude of the negative Goos-H$\ddot{a}$nchen shift is nearly the same as the $\theta_{in}$ with reflectance being one. As a result, the reflected beam mainly consist of beam with negative Goos-H$\ddot{a}$nchen shift. For the transmitted beam, the two parallel beams with positive and negative Goos-H$\ddot{a}$nchen shifts are well separated, as shown in Fig. \ref{figure_GSshiftP}(d).

By extracting the numerical result in Fig. \ref{figure_GSshiftP}(d), the Goos-H$\ddot{a}$nchen shift of the reflected beam is $-108\lambda$. For the transmitted beam, the Goos-H$\ddot{a}$nchen shift of the beam with smaller energy flux is $-195\lambda$, and that of the another beam with larger energy flux is $10\lambda$. Because the FWHM is much smaller than the divergent angle of the incident Gaussian beam, i.e., $\Delta\theta_{in}$, only a small portion of the incident angular spectrum have large negative Goos-H$\ddot{a}$nchen shift. By averaging the Goos-H$\ddot{a}$nchen shift of the whole incident angular spectrum, the negative Goos-H$\ddot{a}$nchen shift of the Gaussian beam is only two-third of the theoretical value of $S_{GH}$ at the resonant peak in Fig. \ref{figure_GSshiftP}(c). If the beam width of the Gaussian beam is further increased, which in turn decreases $\Delta\theta_{in}$, the numerical negative Goos-H$\ddot{a}$nchen shift can further approaches the theoretical value in Fig. \ref{figure_GSshiftP}(c).

\subsection{Sensing of device with small $N_{t}$}

\begin{figure}[tbp]
\scalebox{0.64}{\includegraphics{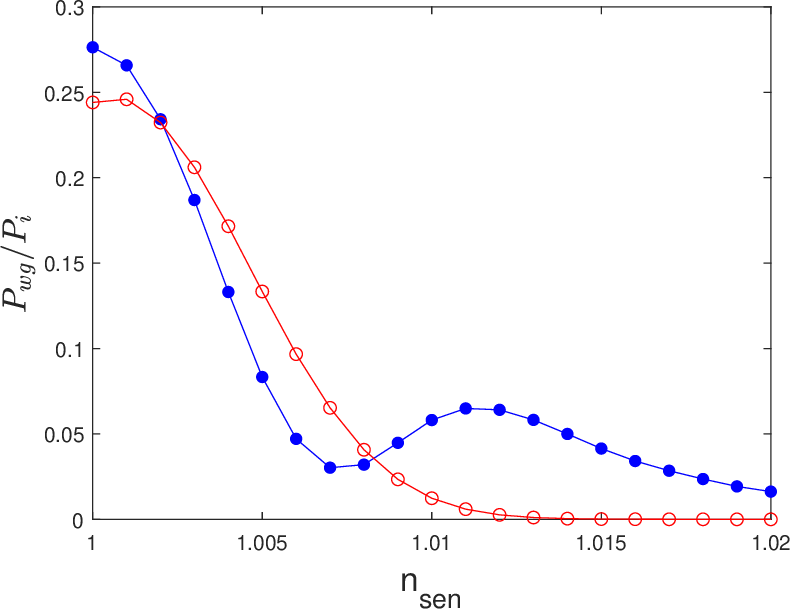}}% Here is how to import EPS art
\caption{ For wavy dielectric grating with $N_{t}=254$ and $N_{t}=307$, $P_{wg}/P_{i}$ under the incidence of s-polarized and p-polarized Gaussian beam, versus $n_{sen}$, are plotted as blue solid and red empty dots, respectively. The parameters of the s-polarized and p-polarized Gaussian beam is the same as those in Figs. \ref{figure_GSshiftS}(d) and \ref{figure_GSshiftP}(d), respectively.  }
\label{figure_GHshiftSense}
\end{figure}

According to the numerical result in Fig. \ref{figure_GSshiftS} and \ref{figure_GSshiftP}, the magnitude of the negative Goos-H$\ddot{a}$nchen shift is larger than the beam width of the Gaussian beam, i.e.,  $w_{0}$. Specifically, the magnitude of the negative Goos-H$\ddot{a}$nchen shift of the transmitted beam, being designated as $S_{GH}^{t}$, is nearly double of $w_{0}$. Thus, at $x=-|S_{GH}^{t}|$, energy flux along $-\hat{x}$ direction within the wavy dielectric grating is sizable; and the magnitude of the incident Gaussian beam is small. If the wavy shape terminates at $x=-|S_{GH}^{t}|$, the termination of the wavy shape has negligible impact on the overlap between the wavy dielectric grating and the incident Gaussian beam. Thus, the energy flux at $x=-|S_{GH}^{t}|$ remains being sizable, so that sizable energy flux could be coupled into the flat dielectric slab, i.e., $P_{wg}$ is sizable. As a result, $P_{wg}$ can be applied as signal for sensing of $n_{sen}$. For the systems in Figs. \ref{figure_GSshiftS}(d) and \ref{figure_GSshiftP}(d), with $N_{t}$ being decreased to be $254$ and $307$, the wavy shape terminates at $x=-N_{t}a=-|S_{GH}^{t}|$ with $S_{GH}^{t}$ being the corresponding negative Goos-H$\ddot{a}$nchen shift of the transmitted beam, respectively. For the s-polarized and p-polarized case, $P_{wg}/P_{i}$ are equal to $0.276$ and $0.244$, respectively, with $P_{i}$ being the power of the incident Gaussian beam. As $n_{sen}$ increases, the parameters $k_{x}$ and frequency of the BIC change, so that the angular spectrum of reflectance and transmittance with a fixed incident wavelength changes. The angular Fano line shape of the quasi-BIC at the incident wavelength shifts away from the incident angle of the Gaussian beam, i.e.,  $\theta_{Inc}$. In the other words, within the range of the angular spectrum of the incident Gaussian beam $[\theta_{Inc}-\Delta\theta_{in},\theta_{Inc}+\Delta\theta_{in}]$, the magnitude of the negative Goos-H$\ddot{a}$nchen shift given by the stationary phase method becomes smaller. Thus, the energy tunneling into the left flat dielectric slab becomes smaller, i.e. $P_{wg}$ becomes smaller. $P_{wg}/P_{i}$ versus $n_{sen}$ are plotted in Fig. \ref{figure_GHshiftSense}.

For s-polarization, $P_{wg}/P_{i}$ monotonically decreases within the range of $n_{sen}\in[1,1.007]$, but increases as $n_{sen}$ further increases. The oscillating behavior of $P_{wg}/P_{i}$ is due to the interfere between the reflected (transmitted) beams with positive and negative Goos-H$\ddot{a}$nchen shifts. The sensitivity of the sensor can be defined as the magnitude of the slope of the function $P_{wg}/P_{i}$ versus $n_{sen}$. The sensitivity of $P_{wg}/P_{i}$ versus $n_{sen}$ can be defined as $\frac{1}{P_{i}}\frac{dP_{wg}}{dn_{air}}$. Within the range of $n_{sen}\in[1.002,1.005]$, $P_{wg}/P_{i}$ is nearly linear function of $n_{sen}$ with sensitivity being $53.83$ RIU$^{-1}$.

For p-polarization, $P_{wg}/P_{i}$ is monotonically decreasing function of $n_{sen}$ within the whole range of $n_{sen}\in[1,1.02]$. In this case, the reflected (transmitted) beams with positive and negative Goos-H$\ddot{a}$nchen shifts are well separated, so that the interference effect is absent. Thus, $P_{wg}/P_{i}$ does not have the oscillating behavior. Within the range of $n_{sen}\in[1.003,1.007]$, $P_{wg}/P_{i}$ is nearly linear function of $n_{sen}$ with sensitivity being $38.20$ RIU$^{-1}$.

In order to extend the sensing range of $n_{sen}$, the parameters of the incident Gaussian beam, i.e., $\lambda$ and $\theta_{Inc}$, can be engineered, so that the quasi-BIC of the systems with varying value of $n_{sen}$ can be excited. For example, if the refractive index of $n_{sen}$ is within the range of $[1.33,1.331]$, one can firstly calculate the wavelength and $k_{x}$ of the cutoff-BIC and accidental-BIC for the system with $n_{sen}=1.33$; secondly, find the parameters $\lambda$ and $\theta_{in}$ of the corresponding quasi-BIC with large negative Goos-H$\ddot{a}$nchen shifts. For the incident Gaussian beam, the incident angle $\theta_{Inc}$ is equal to $\theta_{in}$ with maximum magnitude of the negative Goos-H$\ddot{a}$nchen shift, so that $P_{wg}/P_{i}$ is sizable. As $n_{sen}$ increases from $1.33$ to $1.331$, $P_{wg}/P_{i}$ decreases to zero due to changing of the magnitude of the Goos-H$\ddot{a}$nchen shift. Thus, the sensing range is extended to be $[1.33,1.331]$. In practice, for the systems with varying value of $n_{sen}$, a database of quasi-BICs with negative Goos-H$\ddot{a}$nchen shift being $-N_{t}a$ can be built. The sensor can combine the parameters ($\lambda$ and $\theta_{Inc}$) of the incident Gaussian beam and the measured signal ($P_{wg}/P_{i}$) to infer the value of $n_{sen}$.

\section{Conclusion}

In conclusion, the s-polarized and p-polarized reflectance spectra of wavy dielectric grating are investigated. The cutoff-BIC, which is originated from the coupling between the cutoff waveguide modes and the propagating mode at the light cone, is identified. Under plane wave incidence with wavelength below 500 nm, the sensitivity and FoM of the reflectance spectra to the refractive index of the background medium reach up to $330$ nm/RIU and $3.7\times10^{8}$ RIU$^{-1}$, respectively. The Goos-H$\ddot{a}$nchen shifts of the corresponding quasi-BICs are calculated by applying the stationary phase method to the plane wave incident simulation, and by numerically extracting from the Gaussian beam incident simulation. The magnitude of the negative Goos-H$\ddot{a}$nchen shifts is around $200\lambda$ for the selected quasi-BICs. In the systems with $N_{t}a$ being equal to the magnitude of the numerical negative Goos-H$\ddot{a}$nchen shift, tunneling energy flux to the flat dielectric slab is sensitive to the refractive index of the background medium with high sensitivity. As a result, the system can be applied to build refractive index sensor, which can be integrated with dielectric waveguide in photonic circuit.

\begin{acknowledgments}
This project is supported by the Natural Science Foundation of Guangdong Province of China (Grant No.
2022A1515011578),  the Special Projects in Key Fields of Ordinary Universities in Guangdong Province(New Generation Information Technology, Grant No. 2023ZDZX1007), the Project of Educational Commission of Guangdong Province of China (Grant No. 2021KTSCX064), and the Startup Grant at Guangdong Polytechnic Normal University (Grant No. 2021SDKYA117).
\end{acknowledgments}

\clearpage

\end{document}